\documentclass{JHEP3}

\usepackage{latexsym,amsmath,amssymb}
\pdfoutput=1
\usepackage{graphicx}
\usepackage{slashed}
\usepackage{bm}
\usepackage{epsfig}
\newcommand{\exclude}[1]{}
\usepackage{epsfig}
\usepackage{dcolumn}

\newcommand{\beq}{\begin{equation}}
\newcommand{\eeq}{\end{equation}}
\newcommand{\be}{\begin{eqnarray}}
\newcommand{\ee}{\end{eqnarray}}

\newcommand{\bb}{\bibitem}

\def\dd{ \,\mathrm{d} }

\def\+{\dagger}
\def\la{\langle}
\def\ra{\rangle}
\def\<{\langle}
\def\>{\rangle}

\newcommand{\Lqcd}{\Lambda_{\mathrm{QCD}}}

\title{Large-Scale Magnetic Fields, Dark Energy and QCD.}

\author{Federico R. Urban and Ariel R. Zhitnitsky \\ Department of Physics \& Astronomy, University of British Columbia, Vancouver, B.C. V6T 1Z1, Canada}

\date{\today}

\abstract{Cosmological magnetic fields are being observed with ever increasing correlation lengths, possibly reaching the size of superclusters, therefore disfavouring the conventional picture of generation through primordial seeds later amplified by galaxy-bound dynamo mechanisms.  In this paper we put forward a fundamentally different approach that links such large-scale magnetic fields to the cosmological vacuum energy.  In our scenario the dark energy is due to the Veneziano ghost (which solves the $U(1)_A$ problem in QCD).  The Veneziano ghost couples through the triangle anomaly to the electromagnetic field with a constant which is unambiguously fixed in the standard model.  While this interaction does not produce any physical effects in Minkowski space, it triggers the generation of a magnetic field in an expanding universe at every epoch.  The induced energy of the magnetic field is thus proportional to cosmological vacuum energy: $\rho_{EM}\simeq B^2 \simeq (\frac{\alpha}{4\pi})^2 \rho_{DE}$, $\rho_{DE}$ hence acting as a source for the magnetic energy $\rho_{EM}$.  The corresponding numerical estimate leads to a magnitude in the nG range.  There are two unique and distinctive predictions of our proposal: an uninterrupted active generation of Hubble size correlated magnetic fields throughout the evolution of the universe; the presence of parity violation on the enormous scales $1/H$, which apparently has been already observed in CMB.  These predictions are entirely rooted into the standard model of particle physics.}

\preprint{}
\makeindex

\begin{document}

\section{Introduction}\label{intro}

The origin of cosmological magnetic fields, that is, magnetic fields which permeate the largest structures found in the universe such as galaxies, clusters, and so on, still enjoys the status of ``yet to be solved'' mistery.  Such cosmologically correlated magnetic fields have historically been discovered at ever increasing distances (see the comprehensive reviews~\cite{kronb,grasso,Dolgov:2003xd,Giovannini:2003yn,brand,giov}, mostly due to the fact that experimentalists have come up with new methods and new devices capable of observing their effects in the ever farer and bigger structures in universe.
  
The models and mechanisms that aim at explaining the origin of the large-scale magnetic fields can be roughly filed under two wide categories, astrophysical and cosmological ones.  The astrophysical mechanisms mostly rely on what is typically termed \emph{battery}, see~\cite{grasso,giov} for a review.  Such mechanisms generally provide protogalactic magnetic fields but they are not likely to be correlated much beyond galactic sizes.

On the other hand, cosmological models for the origin of cosmological magnetic fields typically involve some sort of (possibly first order) phase transition, such as the electroweak transition or the QCD confinement one, in order to generate Hubble scale seed fields which then are amplified through the mechanism known as \emph{dynamo}~\cite{brand,giov}.  The typical correlation length of such seeds is typically many orders of magnitude smaller than needed today, unless a process called inverse cascade takes place (see for instance~\cite{ic1,ic2,son}).  An inverse cascade, in brief, spreads highly energetic short wavelength modes to larger correlation lengths at smaller amplitudes.  This effect is expected to take place whenever a helical field evolves in a turbulent plasma, as the early universe mostly is, see again~\cite{grasso,brand,giov} for details.  Depending on the architectures used to model such phenomena, one expects to obtain correlation lengths that vary between a few pc to the very optimistic 100 kpc, but a very weak field intensity.

In the case of an inflationary universe there is also the opportunity of a super-horizon magnetogenesis due to the peculiar expansion character of an inflating universe~\cite{liddle}, which however demands a beyond the standard model (SM) coupling for electromagnetism for the latter is Weyl invariant and is not amplified by gravitational interactions (unless one resorts to the conformal anomaly, see~\cite{sasha}--but that again is inefficient as long as we stick to SM physics).

In all cases, the dynamo is a necessary step in order to revitalise the weakened magnetic fields.  The dynamo must be very efficient for most of the models pushed forward, and, more importantly, should be operating at very large-scales (beyond galaxies).  All these scenarios however incur in a series of difficulties when faced with observations; these problems become profoundly more difficult as observations reach further in intensity and distance, as we argue below.

The structure of our presentation is as follows.  In the following section~\ref{observations} we discuss how observations compare with most theoretical paradigms on the generation of large scale magnetic fields.  We elaborate on a number of problems which appear to be beyond the abilities of such models.  Moving on to section~\ref{out} we offer an alternative approach to resolve the mystery of large scale magnetic field based on the anomalous coupling between dark energy and magnetism.  In subsection~\ref{DE} we review the dynamics of the Veneziano ghost (which solves a fundamental problem in QCD and plays the r\^ole of source for the DE in our framework) in an expanding universe, and sketch how the observed vacuum energy arises, and how the auxiliary conditions on the physical Hilbert space (similar to the Gupta-Bleuler condition in QED) keep the theory unitary.  In subsection~\ref{coupling} we derive the coupling between the Veneziano ghost and electromagnetism via the conventional triangle anomaly.  We shall argue that this coupling, despite being of the same order of magnitude of $g_{\pi^0\gamma\gamma}$ (which describes the $\pi^0\rightarrow\gamma\gamma$ decay), still does not lead to any physical effects in Minkowski space.  By contrast, in an expanding universe when the Veneziano sources the appearance of the cosmological vacuum energy, it does generate large scale magnetic fields which we could be observing now.  In section~\ref{quantit} we outline the arguments leading to our quantitative estimates on such fields, with particular emphasis on the radical differences between the approach pursued here and the far more common generation through primordial seeds.  Lastly, the final section~\ref{end} wraps up the main ideas of the paper with a concise summary with conclusions.

\section{Cosmological Magnetic Fields: Observations vs Theoretical Models}\label{observations}

\subsection{Observations: ever increasing correlation lengths}\label{lengths}

Large-scale magnetic fields have been first discovered in our Milky Way, and subsequently in a number of other galaxies of different sizes and shapes, with characteristic intensity of around a few $\mu$G.  The correlation lengths of such fields range between a kpc and 30 kpc.  However, this is not the end of the story, as magnetic fields of very similar strengths have been observed in clusters of galaxies, where they appear to be correlated over larger distances reaching the Mpc region.  It is important to notice that such fields are not associated with individual galaxies, as they are observed in the intergalactic medium as well~\cite{mpc1,mpc2} (see also the recent papers~\cite{it,ru}).  This poses a serious challenge to astrophysically based mechanisms, for then for some sort of conspiracy a series of unrelated galaxy-bound fields would all choose to align in one specific direction.

As for cosmological magnetogenesis, already at this point one may notice that even the most optimistic and efficient inverse cascades are not going to be able to stretch primordial seeds to such wide lengths, as far as the seeds come from the electroweak phase transition, or even in the case of the QCD one.  In the latter case the best result one can hope to obtain is around 30 kpc (see for instance~\cite{Forbes:2000gr,Forbes:2001vya}) using Son's estimates for inverse cascade parameters~\cite{son}, which are actually already known to be over-optimistic, see~\cite{optim}.  Moreover, recent analyses of the development of magnetic fields in cosmological turbulent plasma show that more realistic parameters for the inverse cascade lead to even smaller correlations~\cite{chiara1,chiara2}.

The most recent hints towards a possible magnetisation of gigantic supercluster structures~\cite{super,hyper}, whose size can easily be two orders of magnitude beyond the clusters within it, although not yet fully conclusive, would represent a further theoretical challenge in modelling magnetogenesis, pushing the correlation lengths further away up to fractions of Gpc.  It is clear that in this case all of the mechanisms mentioned above have very little chance of being successful: such correlation lengths are simply beyond their capabilities.

\subsection{Observations: high redshifts}\label{highz}

In conjunction with the observations of supercluster magnetic fields there are new pieces of evidence that show how $\mu$G magnetic fields were present at much earlier epochs than previously thought~\cite{eearly,early}.  Given this empirical picture, one immediately sees the problem in all mechanism which require a very efficient dynamo amplification be brought to successful completion, for if strong magnetic fields are observed at redshifts as high as $z \sim 5$ there is no time for the galaxies to perform enough rotations and pump up the field as hoped.  In addition to that, it looks at least very unlikely that all galaxies amplify initial, commonly intense seeds just in the same way to return $\mu$G fields everywhere, for that implies roughly the same effective number of turns for all of them.

\subsection{Theoretical models: cosmological seeds?}\label{seeds}

Cosmological seeds suffer from another problem besides those just mentioned, as it has recently been emphasised in~\cite{chiara1}.  This can be schematised as follows.  Primordial magnetic fields trigger the appearance of gravitational waves which would be around during the formation of the light elements (Big Bang Nucleosynthesis - BBN).  Such component is tightly constrained by the beautiful concordance between theory and observations (a review can be found in~\cite{bbn}), and puts problematic limits on the maximal intensity of primordial magnetic fields.  In particular, for non-helical seeds it has been show that neither inflation-born, nor electroweak-born, and not even QCD-born ones would be able to satisfy such limits and provide enough power for a dynamo~\cite{gw1,gw2}.  In the helical case these limits are mitigated, but still on the verge of being excluded, for they demand the most efficient amplification~\cite{chiara1}.

One further issue plaguing inflation-dawned magnetic gemmae is the fact that the back-reaction of the field on the inflaton itself is, in many practical realisations, of non-negligible importance~\cite{slava}.  Indeed, such back-reaction would make inflation unable to reproduce the observed spectrum of temperature fluctuations as observed in the cosmic microwave background (CMB), unless the seeds are smaller that what is required to feed galactic dynamos~\cite{slava}.

Finally, if the dynamo amplification of primordial seeds serves as the main field generation mechanism, it is hard to imagine that low mass dwarf galaxies could produce again just the same $\mu$G fields, because of their relative lower rotation rates.

\subsection{Few more hints on existence of large scale $\mu$G fields }\label{hints}

Briefly, let us list a few scattered ideas supporting an ever existing and all-pervading fairly strong magnetic field.

Coles~\cite{coles} and Kim et al.~\cite{kim} have argued that sufficiently strong magnetic fields should be around during most of the structure formation epoch of the universe, for they would help solving some of the discrepancies between $\Lambda$CDM simulations and the observed structures.

The very simple observation that magnetic fields of intensity spanning at most one order of magnitude in the microgauss are found in completely different environments, from galaxies to intracluster medium, seems to be telling us that such fields are allegedly independent on the matter density they are found to be immersed in.  Indeed, Kronberg~\cite{kronb}, corroborating this suggestion with a number of other pieces of evidence, had proposed that galaxies and clusters have formed in a $\mu$G environmental field, rather than the other way around.

Let us close this overview by mentioning the few models which account for magnetic fields in the formation of very peculiar structures such as filaments, which work better than their magnetic-less counterparts~\cite{fila1,fila2,fila3}, and the correlation between star-burst rates and the intensity of a pervading strong magnetic field within the galaxy, which holds to a good accuracy and may signal how stars are given birth in strongly magnetised surroundings~\cite{totani}.

\section{Coupling Dark Energy with Magnetism}\label{out}

On a completely different side, na\"ively unrelated to large scale magnetic fields, the last decade or so has seen the scientific community steadily realise and acknowledge the existence of the so-called dark energy component which appears to be accounting for over 70\% of the total energy density of the universe~\cite{de1,de2,wmap}.  In this unclustered energy density the entire cosmos is immersed, and its origin is very often thought of as the most intricate problem of modern cosmology and particle physics.

In this work, we propose that the two problems are intertwined at their hearts, and suggest that the same mechanism which explains the fundamental origin of the cosmological vacuum energy should also be able to encompass the generation of cosmologically sized magnetic fields with intensity in the $\mu$G range.  More specifically, it has been recently been shown~\cite{dyn,4d} how dark energy can be explained entirely within the SM, more precisely, within QCD, without the need for any new field or symmetry.  In the proposal~\cite{dyn,4d}, the information about the vacuum dark energy is carried by the so-called Veneziano ghost~\cite{ven}, whose properties in the expanding universe will be reviewed shortly.  The basic idea is that the Veneziano ghost, which is a known, unphysical, degree of freedom in QCD, gives rise non a non-zero vacuum energy (``ghost condensation'' similar to ``gluon condensation" in QCD) if the universe is expanding.  This effect of the Veneziano ghost is in many respects akin to the known phenomenon of particle emission in a time-dependent gravitational background, with the important difference that there is no emission of ``real'', i.e., asymptotic states here\footnote{One should remark here that our preference in using the approach of Veneziano in describing the QCD-related vacuum energy is a matter of convenience.  In principle, the same physics is also (hidden) in Witten's approach to the resolution of the $U(1)$ problem~\cite{witten}.  However, the corresponding technique is much more involved when the system is promoted to time dependent backgrounds, see subsection~\ref{DE} below, and~\cite{dyn} for a detailed discussion.}.  Rather, the effect should be interpreted as the injection of extra energy (in comparison with Minkowski space) into ghost waves when the universe slowly expands.  The average momentum resulting from  this pumping is obviously zero as momentum is still good quantum number in the expanding universe.  Overall, this effect is clearly very tiny as it is proportional to $H$.  What is also important is that the extra energy is stored in the form of ghost waves with momenta $\omega_k \simeq k \simeq H$, as only this much energy can be lent by the expanding background, higher frequencies being exponentially suppressed~\cite{dyn}.  The arguments presented above imply that the typical wavelengths $\lambda_k$ associated with this energy density are of the order of the Hubble parameter, $\lambda_k\sim 1/k\sim 1/H\sim 10\textrm{~Gyr}$, and the corresponding excitations do not clump on scales smaller than this, in contrast with all other types of matter.

The very same ghost field couples via the triangle anomaly to electromagnetism with a constant which is unambiguously fixed in the SM.  While the interaction of the ghost with electromagnetic field is sufficiently strong, it does not produce any physical effects in Minkowski space as a result of the auxiliary conditions on the physical Hilbert space (similar to the Gupta-Bleuler condition in QED) that are necessary to keep the theory unitary.  However, the induced extra time-dependent energy due to the Veneziano ghost in the expanding universe automatically leads to the generation of the physical electromagnetic field.  What is important is that the typical momentum $k^{EM}$ of the generated EM field will be of the same order of magnitude of a typical momentum of the the ghost $k$.  Consequently, typical frequencies of the generated EM field will also be the same order, $\omega_k^{EM} \simeq k^{EM} \simeq H$.

\subsection{Dark energy and the Veneziano ghost}\label{DE}

It has been suggested recently that the solution of cosmological vacuum energy puzzle may not require any new field beyond the SM~\cite{dyn,4d}.  The idea is based on the philosophy that gravitation can not be a truly fundamental interaction, but rather it must be considered as a low-energy effective quantum field theory (QFT)~\cite{eft}.  The first application of this proposal was the computation of the cosmological constant in a spacetime with non-trivial topological structure when a Casimir type effect emerges.  It was shown that the cosmological constant does not vanish if our universe is enclosed in a large but finite manifold with typical size $L \simeq 1/H$, where $H$ is the Hubble parameter.  The cosmological vacuum energy density $\rho_{\Lambda}$ in this framework is expressed in terms of QCD parameters for $N_f=2 $ light flavours as follows~\cite{dyn,4d}:
\be
\label{rhov}
\rho_{\Lambda} \simeq \frac{2N_f  |m_q\la\bar{q}q\ra  |}{m_{\eta'} L} \sim (4.3\cdot 10^{-3} \text{eV})^4 \, .
\ee
This estimate should be compared with the observational value $ \rho_{\Lambda} \approx (2.3\cdot 10^{-3} \text{eV})^4$~\cite{wmap}.  The deviation of the cosmological constant from zero is entirely due to the large but finite size $L$ of the manifold, and, as we have anticipated, should be understood as a Casimir effect in QCD .  This proposal has a very simple and analytically trackable analogue in the 2d Schwinger model~\cite{2d}, and could be tested in upcoming CMB experiments~\cite{cmb}.

The result~(\ref{rhov}) is a direct consequence of the existence of a very special degree of freedom in QCD, that is, the Veneziano ghost~\cite{ven}.  This field is unphysical in Minkowski space (it belongs to the unphysical projection of the Hilbert space) in a way akin to the two unphysical polarisations of the electromagnetic four-potential in conventional QED.  The effective Lagrangian for this field was known already in the '80~\cite{l1,l2}, but it has been recently reworked in a very convenient form for studying its curved spacetime properties as~\cite{dyn}
\be\label{lagC}
{\cal L} &=& \frac{1}{2} D_\mu \hat\phi D^\mu \hat\phi + \frac{1}{2} D_\mu \phi_2 D^\mu \phi_2 - \frac{1}{2} D_\mu \phi_1 D^\mu \phi_1 - \frac{1}{2} m_{\eta'}^2  \hat\phi^2 \nonumber \\
&& + {N_f m_q |\<\bar{q}q\>|}  \cos\left[ \frac{\hat\phi + \phi_2 - \phi_1}{f_{\eta'}} \right] \, ,
\ee
where the covariant derivative $D_\mu$ is defined as $D_\mu = \partial_\mu + \Gamma_\mu$ so that, for instance $D_\mu V^\nu = \partial_\mu V^\nu + \Gamma_{\mu\lambda}^\nu V^\lambda$. The fields appearing in this Lagrangian are
\be\label{names}
\hat\phi = ~\mathrm{physical}~ \eta' \, , \quad \phi_1 = ~\mathrm{ghost} \, , \quad \phi_2 = ~\mathrm{its~partner} \, .
\ee
It is important to realise that the ghost field $\phi_1$ is always paired up with $\phi_2$ in each and every gauge invariant matrix element, as explained in~\cite{dyn}.  The condition that enforces this statement is the Gupta-Bleuler-like condition on the physical Hilbert space ${\cal H}_{\mathrm{phys}}$ for confined QCD, and reads like
\be\label{gb}
(\phi_2 - \phi_1)^{(+)} \left|{\cal H}_{\mathrm{phys}}\right> = 0 \, ,
\ee
where the $(+)$ stands for the positive frequency Fourier components of the quantised fields.  In Minkowski space one can ignore the unphysical  ghost field $\phi_1$ and its partner $\phi_2$ in computing all S matrix elements precisely in the same way as one always ignores the two unphysical photon polarisations when the Gupta-Bleuler condition in QED are imposed. 

However, the requirement~(\ref{gb}) could not be globally satisfied in a general background as explained in details in~\cite{dyn}. This is due to the fact that the Poincar\'e group is no longer a symmetry of a general curved spacetime (including the FLRW universe) and, therefore, it would be not possible to separate positive frequency modes from negative frequency ones in the entire spacetime, in contrast with what happens in Minkowski space where the vector $\partial/\partial t$ is a constant a Killing vector, orthogonal to the $t=\mathrm{const}$ hypersurface, and the corresponding eigenmodes are eigenfunctions of this Killing vector.  The Minkowski separation is maintained throughout the whole space as a consequence of Poincar\'e invariance.  Therefore, all physical effects related to the ghost dynamics are proportional to the deviation from Minkowski spacetime geometry, i.e., to the rate of expansion $H$.  This is the very reason for the Veneziano ghost to exhibit non-trivial dynamics in an expanding universe; gravitational interaction however intervene and change this picture, allowing for the appearance of a non-trivial energy density in the time-dependent background.  We refer the interested reader to the original paper~\cite{dyn} for the details.

One more comment concerns the appearance of the scale $L \simeq 1/H$ in the energy density~(\ref{rhov}).  As it has been extensively explained in the previous letter~\cite{4d}, and also in the longer paper~\cite{dyn}, the potential felt by the ghost and its partner is a Casimir-like energy which is a result of a subtraction procedure that compares the values of the vacuum energy in Minkowski space with that in a general compact manifold (such as a torus of size $L$).  This prescription aims at extracting the physical and measurable portion of the vacuum potential energy of the ghost field, by taking such difference between the vacuum energy in compact curved space and that in infinite Minkowski space.  Essentially, this is our definition of the vacuum energy when the ``renormalised energy density'' is proportional to the departure from Minkowski spacetime geometry and remains finite.  The basic motivation for this definition is the observation (\ref{gb}) that in Minkowski infinite space-time the energy due to the ghost identically vanishes. Technically, it implies that  the Lagrangian itself (\ref{lagC}) does not have any small parameters such as $1/L$.  However, the vacuum energy thereby defined exhibits a Casimir-like effect.  Notice that the correction (which was computed exploiting the topological susceptibility of QCD when the ghost is present) is \emph{linear} in the inverse size of the manifold, not exponentially suppressed, as one would normally expect in the confined phase of QCD where all physical degrees of freedom are massive.

The ghost we are and will be working with here, and whose effects are central for our discussion, was postulated by Veneziano in the context of the $U(1)$ problem.  However, the same problem had been tackled from a different perspective (although in the same large-$N_c$ context) by Witten in~\cite{witten}.  In his approach the ghost field does not ever enter the system, and makes us wonder whether its consequences are physical or just artificious.  Without going into the details (see~\cite{dyn} for a technical explanation), it will be enough here to mention that the curved space effects we have directly computed with the help of, and attributable to, the ghost, are not going to disappear if we follow Witten's approach. The relevant physics will be hidden in the contact term which will depend in a highly non-trivial way on the properties of the spacetime (such as curvature) once the apt renormalisation procedure in the expanding background is performed\footnote{The required procedure, for the non-abelian and strongly interacting theory under consideration is not known yet.}.

\subsection{The ghost-photon anomalous coupling}\label{coupling}
 
Our next step is to include the EM field into the low energy effective Lagrangian~(\ref{lagC}).  First, in order to do so, we need to know the interaction of the $\eta'$ field with photons.  After that, we can recover the interaction of the ghost field $\phi_1$ and its companion $\phi_2$ with the electromagnetic fields because $\phi_1$ and $\phi_2$ always accompany the physical $\eta'$ in a unique way similar to the interaction term~(\ref{lagC}).

The interaction of the $\eta'$ field with photons is a textbook example crafted on the almost identical well known anomalous term describing the $\pi^0\rightarrow \gamma\gamma$ decay, see, e.g.,~\cite{huang}.  The only difference is that $\eta'$ is an isotopical singlet state while $\pi^0$ is an isotopical triplet.  The interaction term is
\be\label{coupeta}
{\cal L}_{\eta'\gamma\gamma} = \frac{\alpha}{8\pi} N_c \sum Q_i^2 \frac{\eta'}{f_{\eta'}} F_{\mu\nu} \tilde F^{\mu\nu} \, .
\ee
where $\alpha$ is the fine-structure constant, $f_{\eta'}$ is the decay constant for the $\eta'$, $N_c$ is the number of colours, and the $Q_i$s are the light quarks electric charges.  Finally, $F_{\mu\nu}$ is the usual electromagnetic field strength (in curved space), and $\tilde F_{\mu\nu} = \epsilon_{\mu\nu\rho\sigma} F^{\mu\nu} /2$ its dual.  We choose $\epsilon^{\mu\nu\rho\sigma} = \epsilon_M^{\mu\nu\rho\sigma} /\sqrt{-g}$ with the Minkowski antisymmetric tensor following from $\epsilon_M^{0123} = +1$, and $g = \det g_{\mu\nu}$ the determinant of the metric tensor.

One should remark here that such kind of anomalous interaction has been studied in great details in particle physics as well as in cosmology.  In particular, the axion (or any other pseudoscalar particle) has exactly the same structure and has been thoroughly analysed in cosmological contexts, see e.g.,~\cite{coup1,coup2,ps1,ps2,coup3,ps3,coup4,ps4,ps5,vanB,vanBibber:2006rb}.

We are not really interested in $\eta'$ physics as the heavy $\eta'$ meson of course is not excited in our universe.  We are interested in the interaction of the ghost field $\phi_1$ and its companion $\phi_2$ with the electromagnetic fields.  The corresponding terms have never been discussed in the literature because they do not appear in any gauge invariant matrix element in Minkowski space as a consequence of the auxiliary condition~(\ref{gb}).  In fact, these unphysical degrees of freedom can be completely integrated out in that case, such that they even disappear at the Lagrangian level (this is exactly the procedure that was adopted in the original paper~\cite{ven}, see also~\cite{dyn} in the context of the present work).  In curved space these fields can not be swept under the carpet as they carry relevant physical consequences: we must explicitly deal with them. 

In order to derive the interaction term between the fields $\phi_1$, $\phi_2$ and $A_\mu$ one should repeat the steps described in~\cite{dyn}.  We can explicitly check that the physical $\eta'$ always enters the Lagrangian in the combination $(\eta' + \phi_2 - \phi_1) / f_{\eta'}$.  Hence, the interaction term we are interested in has the structure
\be\label{coupeta1}
{\cal L}_{(\phi_2-\phi_1)\gamma\gamma} = \frac{\alpha}{8\pi} N_c \sum Q_i^2 \left( \frac{\eta' + \phi_2 - \phi_1}{f_{\eta'}}\right) F_{\mu\nu} \tilde F^{\mu\nu} \, .
\ee
For our future discussions we safely neglect the massive physical $\eta'$ field, and keep  only the ghost field $\phi_1$ and  its companion $\phi_2$ along with the EM field, 
\be\label{lagEM}
{\cal L} &=& - \frac{1}{4} F_{\mu\nu} F^{\mu\nu} + \frac{1}{2} D_\mu \phi_2 D^\mu \phi_2 - \frac{1}{2} D_\mu \phi_1 D^\mu \phi_1 \\
&& - \frac{\alpha}{2\pi} N_c \sum Q_i^2 \left( \frac{\phi_2 - \phi_1}{f_{\eta'}} \right) \vec E \cdot \vec B + {N_f m_q |\<\bar{q}q\>|} \cos\left( \frac{\phi_2 - \phi_1}{f_{\eta'}} \right) \nonumber \, ,
\ee
where the electric and magnetic fields are the usual Minkowski ones (not rescaled by the scale factor of the universe $a(t)$).  We claim that the expression~(\ref{lagEM}) is the exact low energy Lagrangian describing the interaction of the ghost field $\phi_1$ and its companion $\phi_2$ with electromagnetism in the gravitational background defined by the $\Gamma_{\mu}$.  In Minkowski space the expectation value
\be
\label{gb1}
 \left<  {\cal H}_{\mathrm{phys}}| (\phi_2 - \phi_1)|{\cal H}_{\mathrm{phys}}\right> = 0 \, ,
\ee
vanishes, implying that $\phi_1$ and $\phi_2$ are decoupled from QED, as they should in order to preserve the unitarity of the system.

However, as we have pointed out before~\cite{dyn}, the constraint~(\ref{gb}) can not be globally maintained in the entire space in a general curved background.  Thus, the ghost field and its partner do couple to electromagnetism, and consequently we do expect some physical effects to occur as a result of this interaction.  Notice that we are not claiming that the ghost field becomes a propagating degree of freedom, or becomes an asymptotic state.  Rather, we propose the description in terms of the ghost as it is a convenient way to account for the physics hidden in the non-trivial boundary conditions, see~\cite{dyn} once again.  In the context of this discussion, a very illuminating example is that of 2d Rindler spacetime and the associated Unruh effect, see~\cite{eric}, where all different approaches are workable and comparable to give the same result.  One should emphasise that the Veneziano ghost is very different from all other types of ghosts, including the conventional Fadeev-Popov fields.  The peculiar feature of the Veneziano ghost resides in its close connection with the topological properties of the theory, and expresses the necessity to sum over different topological sectors in QCD~\cite{eric}.  This uniqueness manifests itself, in particular, in the spectrum of the Veneziano ghost: while conventional ghosts may have arbitrary large frequencies, and essentially, are introduced only to cancel unphysical polarisations of gauge fields for arbitrary large $\omega$, typical frequencies of the Veneziano ghosts are order of the horizon scale, $\omega\sim H$.

The most immediate consequence of the interaction term~(\ref{coupeta1}) is that the magnetic field which will be generated this coupling will have a typical Fourier mode $k^{EM}\simeq k$ of the same order of magnitude of the ghost mode.  On the other hand, gravity can only lend energies of order $\omega_k \simeq k \simeq H$, higher frequencies being exponentially suppressed.  Consequently, a typical EM mode will be around $\omega_k^{EM} \simeq k^{EM} \simeq H$.  Moreover, this interaction is active at every epoch, and we shall see in the forthcoming section how~(\ref{lagEM}) naturally explains the $\mu$G intensity apparently observed in galaxies and clusters, and maybe needed from the very beginning of structure formation.  To conclude, let us notice once again that everything we have been discussing so far is part of the SM, and the coupling constants are all known.  However, all our conclusions will apply to most (pseudo)scalar models of dark energy, as long as they are augmented with a coupling with the same structure as~(\ref{coupeta1}); in this sense the results we will discuss in the upcoming section are general, and widely applicable to dark energy theories.

\section{Dark Energy and Large Scale Magnetic Field, or Two Sides of the Same Coin}\label{quantit}

The Lagrangian density~(\ref{lagEM}) explicitly contains a coupling between the ghost and the $P$-odd operator $\vec E \cdot \vec B$ made of electromagnetic fields.  The structure of this term is identical to the textbook example describing the anomalous $\pi^0\rightarrow \gamma\gamma$  decay.  Let us explore the consequences of this interaction in our specific circumstances.  While the structure of the ghost-photon interaction is a photocopy of that of the pion-photon one, there is a fundamental difference between the two: $\pi^0$ is a massive physical particle which can decay to two photons, whereas the ghost is massless and unphysical, and can not decay into two photons.  The ghost field in an expanding universe should be treated as the large correlated classical field which emerges from a non-zero expectation value $\left<  {\cal H}_{\mathrm{phys}}| (\phi_2 - \phi_1)|{\cal H}_{\mathrm{phys}}\right> \neq 0$ as explained in~\cite{dyn}.  The Fourier expansion of these classical fields $\phi_2$ and $ \phi_1$ is saturated by very low frequencies $\omega_k \simeq k \simeq H$, while higher frequency modes  $\omega_k \gg H$ are strongly suppressed as a result of the relative suppression of the so-called Bogolubov coefficients~\cite{dyn}. 

\subsection{Fast equilibration: first estimates of $B$}

For future convenience we introduce the dimensionless coupling constant which appears in our basic expression~(\ref{lagEM})
\be\label{beta}
\beta \equiv \frac{\alpha}{2\pi} N_c \sum Q_i^2 \, .
\ee

In nature $\beta \ll 1$, but we still want to study the rate of energy transfer from dark energy (which is represented by the ghost field) to the electromagnetic energy as a function of $\beta$.  Hence, we treat $\beta$ as a free parameter in this section.  If the fields are interacting sufficiently strongly ($\beta\gg 1$) and the potential minimum is reached sufficiently rapidly, one can estimate the expectation value $\< \vec E \cdot \vec B\> $ in terms of the external background field $\<\phi_2 - \phi_1\>$, which is treated as a source.  By following this procedure we arrive to
\be\label{minim}
 \langle \vec E \cdot \vec B \rangle =  \frac{1}{\beta}{N_f m_q |\<\bar{q}q\>|} \langle \frac{\phi_2 - \phi_1}{f_{\eta'}} \rangle \simeq  \frac{1}{\beta}\Lqcd^3H \, , ~~ \beta\gg 1 \, .
\ee

As expected, in Minkowski space there will be no generation of EM field as $\<\phi_2 - \phi_1\>=0$ according to~(\ref{gb}),~(\ref{gb1}).  In an expanding universe $\<\phi_2 - \phi_1\>$ is proportional to the deviation from Minkowski space, and is expected to be around $H$.  In this case the magnetic field is produced as an outcome of the energy flow from the ghost to the EM field.  At the time at which each field mode is born, our equations are symmetric under the permutation of $\vec E$ and $\vec B$.  Therefore, one could estimate the absolute value of $|\vec{B}|$ as\footnote{The exact statement is $ [\<\vec E^2\>+ \< \vec B^2\> ] \geq 2 \<\vec E \cdot \vec B\> $ but the electric field will be screened soon after it is generated, and it can then be neglected in the evolution.}, $ \<\vec E \cdot \vec B\>\simeq \<\vec B^2\> $. Of course, the evolution of the electric field and magnetic fields are drastically different  as electric charges do exist in nature while magnetic ones do not.  However, we expect that $\<\vec E \cdot \vec B\>\simeq \<\vec B^2\>$ is a reasonable approximation for the absolute value of $|\vec{B}|$ at its birth.  Consequently, a simple estimate of the intensity of the magnetic field based on the assumption that the system (ghost + EM) can reach its minimum energy configuration sufficiently quickly (fast equilibration) can be presented as
\be\label{wrong}
\<\vec B^2\>\simeq  \langle \vec E \cdot \vec B \rangle \simeq  \frac{\Lqcd^3H}{\beta} \simeq \frac{\rho_{DE}}{\beta} \, , ~~ \beta\gg 1\, ,
\ee
where $\rho_{DE} \simeq \Lqcd^3H$ in our framework.  For large coupling constant $\beta$ this estimate is  totally justified.  Indeed, eq.~(\ref{minim}) corresponds to the minimum of the potential energy of the system\footnote{In particular, a similar procedure of minimisation of the effective potential allows to compute the exact vacuum expectation value for the gluon field $\<\frac{ \alpha_{s}}{8 \pi} G_{\mu \nu}^{a}\tilde{G}^{\mu \nu a} \> =\frac{\theta}{N_f} m_q \<\bar{q}q\> $ which is analogous to (\ref{minim}) when  the colour gluon fields $G^{\mu \nu a} $ 
replace the EM fields, and the so-called $\theta$ parameter of QCD replaces  the expectation value of the ghost field  $\<\phi_2 - \phi_1\>$.  The equilibrium for $\<\frac{ \alpha_{s}}{8 \pi} G_{\mu \nu}^{a}\tilde{G}^{\mu \nu a} \>$ is obviously achieved very quickly and the relation
$\<\frac{ \alpha_{s}}{8 \pi} G_{\mu \nu}^{a}\tilde{G}^{\mu \nu a} \> =\frac{\theta}{N_f} m_q \<\bar{q}q\> $ becomes exact at the minimum of the potential.  In fact, one can differentiate this relation with respect to $\theta$ one more time to arrive to well-known exact Ward Identity for the topological susceptibility, $i\int dx \<\frac{ \alpha_{s}}{8 \pi} G_{\mu \nu}^{a}\tilde{G}^{\mu \nu a} (x), \frac{ \alpha_{s}}{8 \pi} G_{\mu \nu}^{a}\tilde{G}^{\mu \nu a}(0) \>=\frac{m_q}{N_f} \<\bar{q}q\> $, see, e.g.,~\cite{Halperin:1998rc} and references therein.}, and it becomes a precise statement  for $\beta\gg 1$ when the equilibration is achieved faster than a Hubble time $1/H$.
  
However, in nature $\beta\ll 1$. Let us see what is happening with our formula~(\ref{wrong}) when we start to decrease $\beta$.  This equation tells us that we generate magnetic energy which (parametrically!) starts to exceed the energy of the source when $\beta\sim 1$.  Clearly this can not be true, and the loophole in the above line of arguments lies in the assumption of equilibrium.  The minimisation procedure described above works only when the reactions involved in transferring energy from one source (the ghost) to a recipient (the electromagnetic field) are efficient enough to be in equilibrium.  The lowest energy of the system simply can not be achieved when the coupling between the two components is weak.

Thus, in a slowly expanding universe one needs to have a handle on the rate of energy transfer at each epoch, and compare it with the Hubble time, just as it is typically done in studying the thermal history of particle species in the early universe.  The difference here is that this rate depends itself on the rate of expansion $H$ due to the fact that the effective coupling constant is proportional to $\<\phi_2 - \phi_1\>\simeq H$ as formula~(\ref{minim}) states.  As we will see shortly, in one Hubble time only a very small fraction of the ghost's potential energy is transferred to the magnetic field at $\beta\ll 1$, thereby mining the foundations of the minimisation procedure.  Yet, there is an important lesson to be learnt from this discussion: the dynamics described by the Lagrangian~(\ref{lagEM}) does lead to energy transfer from the ghost field (which is the DE in our  model) into the magnetic field, though numerically formula~(\ref{wrong}) can not be trusted for the physically relevant case $\beta\ll 1$. 

\subsection{The time scales} 

As we mentioned above, in order to understand the dynamical effects of the coupling between the ghost and electromagnetism one should  look for the rate at which energy is transferred while the universe expands.  In order to do so let us write down the Hamiltonian for the system~(\ref{lagEM}) as
\be\label{ham}
\mathrm{H} &=& \frac{1}{2} \left( B^2 + E^2 \right) + \frac{1}{2} D_\mu \phi_2 D^\mu \phi_2 - \frac{1}{2} D_\mu \phi_1 D^\mu \phi_1 \\
&& + \frac{\alpha}{2\pi} N_c \sum Q_i^2 \left( \frac{\phi_2 - \phi_1}{f_{\eta'}} \right) \vec E \cdot \vec B - {N_f m_q |\<\bar{q}q\>|}  \cos\left( \frac{\phi_2 - \phi_1}{f_{\eta'}} \right) \nonumber \, .
\ee
As we discussed earlier in the text, neither the Lagrangian~(\ref{lagEM}), nor the Hamiltonian~(\ref{ham}) contains any small coupling constant in their definitions as all the parameters describing the system are known SM parameters.  All small effects are proportional to $H/\Lqcd\sim 10^{-41}$ and are brought about only at the level of  the ``renormalisation procedure'', i.e., when the subtraction is explicitly performed.  If we are computing the energy related to the magnetic field in Minkowski space, this identically vanishes~(\ref{minim}) as a consequence of the conditions~(\ref{gb}).

The time derivative $\dd\mathrm{H}/\dd t $ governs the efficiency of the energy flow between ghost and electromagnetic fields.  Explicitly, $\dd\mathrm{H}/\dd t =  0$ implies that
\be\label{hamt}
\frac{\dd}{\dd t} \left[ \frac{1}{2} \left( B^2 + E^2 \right)\right] \simeq - \beta \left( \frac{\dot\phi_2 - \dot\phi_1}{f_{\eta'}} \right) \vec E \cdot \vec B - \beta \left( \frac{\phi_2 - \phi_1}{f_{\eta'}} \right) \frac{\dd}{\dd t} \left[ \vec E \cdot \vec B \right] - \frac{\dd}{\dd t}\mathrm{{H}_{ghost}} \, ,
\ee
where $\dd\mathrm{{H}_{ghost}} / \dd t$ essentially describes the dynamics of the dark energy component, and can be neglected in this simple evaluation.
 
We want to gain some intuition about the rate of the energy transfer by considering the case of small $\beta$ when we know that the minimisation procedure is not justified.  Assuming for simplicity that the Hubble parameter is a constant and the rate of energy transfer is also a constant, one can estimate the typical time (relaxation period) $\tau_0$ which is required for the system to reach its equilibrium value.  Indeed, the left hand side of eq.~(\ref{hamt}) can be approximated as
 \be\label{tau1}
\text{l.h.s. of eq.~(\ref{hamt}) }  =   \frac{\dd}{\dd t} \left[ \frac{1}{2}\la  B^2 + E^2 \ra\right] \simeq \frac{\rho_{EM}}{\tau_0} \, .
\ee
At the same time, the right hand side of~(\ref{hamt}) is
\be\label{tau2}
\text{r.h.s. of eq.~(\ref{hamt}) } \simeq - \beta \la \frac{\dot\phi_2 - \dot\phi_1}{f_{\eta'}} \ra \rho_{EM} - \la \frac{\phi_2 - \phi_1}{f_{\eta'}} \ra \frac{\rho_{EM}}{\tau_0} \, ,
\ee
where we have substituted $\la\vec E \cdot \vec B\ra \simeq  \rho_{EM}$ and its time derivative with a $1/\tau_0$.  The last step we need to do is simply to plug in the values $ \la \dot\phi_2 - \dot\phi_1 \ra / f_{\eta'} \simeq \la \phi_2 - \phi_1 \ra \simeq H$ (non-zero solely due to the expanding gravitational background).  A more precise numerical estimate for $\la \dot\phi_2 - \dot\phi_1\ra$ will be given below.  Comparing~(\ref{tau1}) with~(\ref{tau2}) one can immediately infer that
\be\label{tau}
\tau_0\sim \frac{1}{\beta H} \, , ~~~~ \beta\ll 1 \, ,
\ee
which is the main result of this subsection.

\subsection{Slow equilibration: more realistic values for $B$}

Now we want to see what happens with our estimate (\ref{wrong}) when $\beta$ decreases, as it corresponds to the physical value~(\ref{tau}).  From what we have said above, it is clear that the time scale which is required to attain equilibrium will be order of the Hubble time $\tau_0\simeq 1/H$ for $\beta\sim 1$ when the minimisation formulae can still be marginally trusted.  In this case, a finite portion of DE of order one can be transferred into the magnetic energy during a Hubble time.  When $\beta$ decreases even further to reach the physically interesting region $\beta\ll 1$ the formula~(\ref{tau}) suggests that in order to reach equilibrium one needs a time which is $\beta^{-1}\gg 1$ times larger than the Hubble time $1/H$, which really makes no sense.  The appropriate interpretation in this case is that equilibrium will be never achieved with such small coupling constant during one Hubble time.  Instead, a very small portion $\beta$ of the available energy can be at most injected into the magnetic field within the same Hubble time.  

This is however not the end of the story for small $\beta\ll 1$.  The point is that, for $\beta\sim 1$, each event leading to equilibration transfers an amount of energy of order one.  For small $\beta\ll 1$ this argument does not hold anymore: not only the equilibration time increases as a result of fewer events per unit time, see eq.~(\ref{tau2}), also a smaller portion of the available energy will be transferred to the EM field per collision.  It follows that a more realistic estimate shall account for an additional suppression factor $\beta$ in comparison with case $\beta\sim 1$, when formula~(\ref{wrong}) can be still marginally trusted.

With this interpretation in mind we arrive at our final estimate for the magnetic energy that has flowed from the ghost field
\be\label{final}
\rho_{EM}  \simeq \beta^2 \la \frac{\dot\phi_2 - \dot\phi_1}{ H f_{\eta'}} \ra \rho_{DE}  \simeq
\left(\frac{\alpha}{2\pi} N_c \sum Q_i^2\right)^2  \la \frac{\dot\phi_2 - \dot\phi_1}{ H f_{\eta'}} \ra \rho_{DE} \, ,
\ee
where we inserted the extra small parameter $\beta^2$ accounting for the two suppression factors mentioned above, accounting for the fact that  only a small fraction of the total available energy will effectively exchanged during one Hubble time.

One should remark here that the $\beta^2$ suppression which enters in~(\ref{final}) can be intuitively understood by considering a system of particles with magnetic moment.  In this case, as is known, the interaction between the magnetic moment and an external magnetic field is proportional to the coupling constant $e$.  However, the contribution to the energy due to the induced magnetic moment is proportional to $e^2$.  In different words, the magnetic susceptibility of the system goes as $e^2$ rather than linearly with $e$, in spite of the fact that the strength of the interaction itself is $\propto e$.  Our analysis of the induced EM field follows the footprints of the analogous induced magnetic argument, where now the r\^ole of the coupling constant $e$ is played by the parameter $\beta$, and where the source of the energy is $\rho_{DE}$ rather than the external magnetic field.  Finally, in our expressions above we assumed that the Hubble constant does not depend on time, so that the transfer rate is time-independent itself.

A few comments are in order before we proceed with the numerical evaluations.  First of all, eq.~(\ref{final}) has the following parametrical dependence on fine structure constant $\alpha$ and Hubble constant $H$:
\be\label{final1}
\rho_{EM}  \simeq  \la \vec{B}^2\ra \simeq \left(\frac{\alpha}{2\pi}\right)^2  H \Lqcd^3 , ~~~ {\rm where} ~~  \rho_{DE}  \simeq H \Lqcd^3 \, .
\ee
In this form it is easy to interpret the appearance of each parameter.  Dark energy is a QCD effect related to the mismatch between vacuum energies in infinite Minkowski and compact or curved spaces, and it is conceivably proportional to the rate of the expansion of the universe.  As is known, the EM field does not interact directly with gravity, however, it does interact (through a standard triangle anomaly) with a (pseudo) scalar ghost field which gives rise to a correlated external source for the EM field in an expanding universe.  This explains the extra parameter $(\alpha/2\pi)^2$ in~(\ref{final1}).  There are other numerical factors of order one which would also appear in~(\ref{final1}), which for now are glossed over to simplify the interpretation.  However, we believe that our formula~(\ref{final1}) has a well understood and physically motivated behavior in terms of the relevant physical parameters $\alpha$, $H$, $\Lqcd$.

We are now in the position to make a more quantitative estimate for the magnetic field as the combination which enters expression~(\ref{final}) and which describes the dynamics of the ghost field, that is, $c\sim \la\dot\phi_2 - \dot\phi_1\ra / (H f_{\eta'})$, had been previously discussed in~\cite{dyn} and it essentially corresponds to the initial field velocity in the classical potential.  An uncertainty in value of this constant, $c$, is not related to any physics beyond the SM, but is simply determined by the initial conditions when the dynamics of $\phi_2$ and $\phi_1$ are treated classically\footnote{The classical treatment of the system is by far the widest chosen approach in dealing with dark energy matters.  However, while in other, perhaps more familiar, cosmological models such as inflation, the passage from quantum to classical is justified \emph{a posteriori} (see for instance the discussion in~\cite{liddle}, section 7.4.7, and reference to original works therein), in coping with our quantum fields we do not expect such a ``little miracle'' to happen.  In other words, the quantum nature of our fields which appears in their non-trivial dependence on the gravitational background as well as on the global properties of the manifold, is brought in at the level of the renormalisation procedure, and is therefore not describable in a purely classical approach.  Nevertheless, as long as we are interested in order of magnitude estimates, it will suffice to confine ourselves within the boundaries of such classical framework.}.  For our order of magnitude estimate the specific value of this parameter is unimportant, and, following~\cite{dyn}, we will fix it at the value of $c \sim 10^{-3}$.

With these remarks in mind we can finally work out some numerology.  The typical value for the energy density of the magnetic field which is generated by the Veneziano ghost, which in turn is responsible for the cosmological dark energy, is
\be\label{final2}
B^2 &\simeq& \left( \frac{\alpha}{2\pi} N_c \sum Q_i^2 \right)^2\cdot c \cdot \rho_{DE} \, , ~~~~~ c \sim 10^{-3} \, .
\ee
From this expression it straightforwardly follows that during the last Hubble time of life of the universe, an ${\cal O}(H_0^{-1})$ correlated magnetic field is born with intensity
\be\label{numeric}
B&\simeq &  \frac{\alpha}{2\pi} \sqrt{c }\cdot (2.3\cdot 10^{-3} \text{eV})^2\sim \text{nG} \, , ~~~~ 1 \text{G}= 1.95\cdot 10^{-2}  (\text{eV})^2 \, .
\ee
As we mentioned previously, formula~(\ref{numeric}) should be treated as an order of magnitude estimate due to a number of numerical factors which have been consistently neglected while arriving at our final analytical result~(\ref{final}) and its numerical expression~(\ref{numeric}).  We should emphasise that the most important outcome here is not the precise numerical value found in~(\ref{numeric}), which is likely to change due to the very complicated evolution of the field in its most recent history ($z \sim 1$) well after its formation.  Rather, the main result of this paper is the observation that due to the coupling with DE the EM field will be correlated on enormous scales of order $1/H$.  This distinguishes our mechanism from everything which has been suggested previously.  Another important qualitative consequence is the prediction that parity will be locally violated on the same scales for which the field does not change, that is, $1/H$ as a result of the pseudoscalar nature of the DE field.  This distinct and unambigious prediction can be tested in the CMB sky, where apparently parity violation on such scales indeed has been observed, see~\cite{Kim:2010gf,Kim:2010gd} (we will comment further on this aspect below).  Finally, our field is highly helical with typical wavelengths $\lambda_k\sim 1/H$; therefore, the induced $\vec B$ is expected to flip sign on the same scale.  We notice in passing that a $\text{nG}$ intensity would account for the observed galactic field if it were frozen in the pre-galactic plasma.  In spite of its attractiveness however, this possibility should be taken with a grain of salt; furthermore, the analysis of the evolution of such fields is beyond the scope of the present work.

This mechanism operates all the time, and there will be an uninterrupted flow of magnetic fields produced at different correlation lengths (proportional to the Hubble parameter).  Such fields however, being proportional to the vacuum energy component, will be of significantly relative weaker amplitude compared to the other components of the universe, following the twin evolution of the dark energy.  Nevertheless, they could still behave as seeds for plasma mechanisms to process them, the outcome of which is beyond the scope of this paper and shall not be pursued any further.

\subsection{Some applications}

As we have already mentioned, interactions of the form~(\ref{coupeta1}) have been detailedly exploited and analysed in the literature, initially in the context of the anomalous axion-photon interaction, and then extended to general pseudoscalar interactions.  What makes this work depart from all the paper referenced in the bibliography is that we are dealing entirely with SM physics, including the interaction~(\ref{coupeta1}).  We should remark here that this interaction does not violate $P$ and $CP$ invariance on the fundamental level, similarly to the $\pi^0\rightarrow 2 \gamma$ decay.  However, on small scales, one can interpret the interaction~(\ref{coupeta1}) as a $P$ and $CP$ violating coupling similar to the $\theta$ term in QCD.  Such a violation occurs only locally, on the correlation scales $\lambda_k\sim 1/H$, while globally one should expect $P$ and $CP$ conservation according to the symmetry of the fundamental interactions~(\ref{lagEM}).

In this paper we have worked out the first application of such interaction in our model, that is, the generation of cosmological magnetic fields.  It is easy to see, however, that there are several possible effects in addition to what has just been computed, especially in connection with possible observables which would be able to confirm or falsify our framework.  We will limit ourselves to a mention for some of these possibilities, the specific details of which are beyond the scope of the present work and will be addressed in future publications.

\begin{itemize}
   \item Early universe large scale magnetic fields have a sound impact on the mechanisms of structure formation (especially an active source of nanogauss intensity); there exists a vast literature on the subject, although this point is still somewhat overlooked: we refer to the comprehensive reviews~\cite{kronb} and~\cite{grasso} and to section~\ref{hints} where some interesting consequences of primordial magnetism in the formation of early structures are investigated.
   \item The presence of a pre-decoupling magnetic field is able to leave significant imprints in the cosmic microwave background radiation (CMB).  The literature on this specific topic is particularly copious, and we shall mention only a few specific papers.  The effects of magnetic helicity (particularly relevant for us since in our case the magnetic field is highly helical) have been discussed in~\cite{hel1,hel2}.  Primordial magnetism is also able to de-polarise the CMB (which acquired a small degree of linear polarisation due to the finite thickness of the last scattering surface), see~\cite{pol1,pol2,pol3,pol4,pol5}.  Finally, an impressive and complete analysis of the distortions caused by pre-decoupling magnetic fields (stochastically distributed) to the CMB acoustic peaks has been recently undertaken in the series of papers~\cite{gk1,gk2,gk3,gk4,gk5}.  The novel aspect of our proposal compared to the models adopted in all the aforementioned efforts is that our source is active at all times, has a highly non-trivial redshift dependence, and is capable of producing otherwise forbidden by parity cross-correlation spectra such as those known as $E B$ and $T B$ on top of the distortions predicted for the usual $TT$ and $EE$ (and $BB$) autocorrelations and the parity even $TE$ cross correlation.  Anomalous parity violation in the low multipoles of the CMB has been discussed in~\cite{Kim:2010gf,Kim:2010gd}.
   \item In addition to the pervading magnetic fields generated through the Veneziano ghost, one could also look for other signatures of the parity-violating\footnote{$P$ and $CP$ violation may occur locally, not globally, as explained above.} coupling~(\ref{coupeta1}); in the early expansion history this interaction is able to trigger a sizeable amount of birefringence of the CMB~\cite{bir1,bir2,bir3} (the peculiar effects of the parity violation have been considered in~\cite{pv1,pv2,pv3,pv4}); at late times such coupling can be sought after in its impact to the travelling light signals from the CMB, or any emitting source in the universe, to us, the observers, see again~\cite{coup2,coup3,coup4} and~\cite{jain1,jain2,jain3}.  The latter is of prominent relevance due to the recent claims of a preferred orientation of the polarisation angles of quasars in optical wavelengths~\cite{hut1,hut2,hut3} which seems to disappear at radio frequencies~\cite{pa1} (see also~\cite{pa2} for an analysis of the possible biases affecting these results): a possible explanation is again to be found in the combined effect of a background magnetic field and an anomalous coupling of~(\ref{coupeta1}) fame which mixes light and the pseudoscalar field~\cite{jain1}.
\end{itemize}

\section{Results and conclusion}\label{end}

In this work we have proposed that the origin of the observed cosmological magnetic fields, correlated over scales that stretch from galaxies to superclusters, is tied to the existence and evolution of the vacuum energy component of the universe.  

Here we have shown how the very same field which provides the dark energy in our framework~\cite{dyn}, together with its Gupta-Bleuler partner, possesses an anomalous coupling (the ordinary triangle anomaly) to the magnetic field and can trigger the generation of magnetic fields of the observed intensity with correlation lengths of order of the Hubble parameter today.  This interaction is present at all times starting from the QCD phase transition, and is continuously producing magnetic fields, whose strength is consequently linked to that of the vacuum energy, times a dimensionless coefficient of about $(\alpha / 4\pi)^2$ due to the weakness of the electromagnetic interactions.  This combination, $\alpha^2 \rho_{DE} / 16\pi^2$, where in our set up $\rho_{DE} \simeq H \Lqcd^3$, is precisely the magnetic field strength (squared) one needs to account for the observational evidence.

The list of intricacies most large-scale magnetic field models have to face does not clash with the scheme we have outlined.  Correlation lengths up to the size of the universe at each epoch are a direct consequence of the properties of the Veneziano ghost.  Magnetic fields will be generated at all times, even at higher redshifts, though their magnitudes will experience a standard $1/a^2(t)$ suppression due to the expansion of the universe.  There is no need for seeds to feed a dynamo with, as the largest scales are generated last, and with already the correct strength, without the risk of overwhelming BBN with gravitational waves.

Distinctive predictions and unique features of our proposal are the generation of the magnetic field with magnitude in the nG range at all scales up to the Hubble radius today, and its grounds to be found entirely within familiar SM physics, without the need for any new fields, unconventional couplings to gravity or any modification of gravity itself.  This mechanism is also unique in predicting parity violation on the largest scales $1/H$, which apparently is already supported by CMB observations~\cite{Kim:2010gf, Kim:2010gd}.  If these predictions were to be confirmed by future PLANCK observations, it would be a strong (but still implicit) hint towards the QCD nature of dark energy, together with its fundamental kinship with large scale magnetic fields, in the way detailed in the present work.  

Our final comment concerns the scales involved in the model. Our proposal leads to the order of magnitude estimate $ B \simeq \frac{\alpha}{2\pi} \sqrt{H \Lqcd^3}\sim$ nG, which is approximately the value that, by simple adiabatic compression, could explain the field observed at all scales, from galaxies to superclusters.  It is already the second time that this ``accidental coincidence'' happens, the first one being the dark energy density itself, written as $\rho_{DE}  \simeq H \Lqcd^3 \sim (10^{-3} \text{eV})^4$.  We consider this ``coincidence'' as encouraging support for the entire framework.  The SM coupling of the EM field with the DE field leads to the observed $\mu$G galactic magnetic field (assuming it were frozen in the pre-galactic plasma).  It is difficult to invent another scheme when these relations hold, see, e.g.,~\cite{Carroll:1998zi}.  This ``coincidental'' relation should be added to our ``Fine-tuning without fine-tuning'' section from~\cite{dyn} where we clarify how the intimidating list of fine-tuning issues which always plagues dark energy models, possesses simple explanations without the need to introduce new fields, which come with new interactions, new coupling constants, and new symmetries.  Instead, all our results are based on the paradigm according to which the ``physical dark energy'' is related to the deviation of the vacuum energy from infinite Minkowski space similarly to the Casimir effect, while all SM fields couple to the dark energy field (the Veneziano ghost) in the well understood way dictated univocally by the standard model of particle physics.

\section*{Acknowledgements}
AZ thanks P Naselsky for discussions on the possibilities to test some of the ideas formulated in this paper (such as Hubble scale magnetic fields and parity violation on the same scales) in the data being gathered and analysed by the PLANCK collaboration.  FU would like to thank S Habib for valuable conversations.  This research was supported in part by the Natural Sciences and Engineering Research Council of Canada.


\begin{thebibliography}{99}

\bb{kronb}
P.~P.~Kronberg,
  %``Extragalactic magnetic fields,''
  Rept.\ Prog.\ Phys.\  {\bf 57}, 325 (1994).
  %%CITATION = RPPHA,57,325;%%

\bb{grasso}
D.~Grasso and H.~R.~Rubinstein,
  %``Magnetic fields in the early universe,''
  Phys.\ Rept.\  {\bf 348}, 163 (2001)
  [arXiv:astro-ph/0009061].
  %%CITATION = PRPLC,348,163;%%

\bb{Dolgov:2003xd}
  A.~D.~Dolgov,
  %``Magnetic fields in cosmology,''
  arXiv:astro-ph/0306443.
  %%CITATION = ASTRO-PH/0306443;%%
 
\bb{Giovannini:2003yn}
  M.~Giovannini,
  %``The magnetized universe,''
  Int.\ J.\ Mod.\ Phys.\  D {\bf 13}, 391 (2004)
  [arXiv:astro-ph/0312614].
  %%CITATION = IMPAE,D13,391;%%
  
\bb{brand}
A.~Brandenburg and K.~Subramanian,
  %``Astrophysical magnetic fields and nonlinear dynamo theory,''
  Phys.\ Rept.\  {\bf 417}, 1 (2005)
  [arXiv:astro-ph/0405052].
  %%CITATION = PRPLC,417,1;%%

\bb{giov}
M.~Giovannini,
  %``Magnetic fields, strings and cosmology,''
  Lect.\ Notes Phys.\  {\bf 737}, 863 (2008)
  [arXiv:astro-ph/0612378].
  %%CITATION = LNPHA,737,863;%%

\bb{ic1}
A.~Brandenburg, K.~Enqvist and P.~Olesen,
  %``Large-scale magnetic fields from hydromagnetic turbulence in the very early
  %universe,''
  Phys.\ Rev.\  D {\bf 54}, 1291 (1996)
  [arXiv:astro-ph/9602031].
  %%CITATION = PHRVA,D54,1291;%%

\bb{ic2}
A.~Brandenburg, K.~Enqvist and P.~Olesen,
  %``The effect of Silk damping on primordial magnetic fields,''
  Phys.\ Lett.\  B {\bf 392}, 395 (1997)
  [arXiv:hep-ph/9608422].
  %%CITATION = PHLTA,B392,395;%%

\bb{son}
D.~T.~Son,
  %``Magnetohydrodynamics of the early universe and the evolution of  primordial
  %magnetic fields,''
  Phys.\ Rev.\  D {\bf 59}, 063008 (1999)
  [arXiv:hep-ph/9803412].
  %%CITATION = PHRVA,D59,063008;%%

\bb{liddle}
A.~R.~Liddle and D.~H.~Lyth,
  ``The Primordial Density Perturbation: Cosmology, Inflation and the Origin of Structure,'' {\it Cambridge University Press; Revised edition}, Cambridge, UK (2009)
  %%CITATION = ISBN-13-9780521828499;%%

\bb{sasha}
A.~Dolgov,
  %``Breaking Of Conformal Invariance And Electromagnetic Field Generation In
  %The Universe,''
  Phys.\ Rev.\  D {\bf 48}, 2499 (1993)
  [arXiv:hep-ph/9301280].
  %%CITATION = PHRVA,D48,2499;%%

\bb{mpc1}
B.~M.~Gaensler, R.~Beck and L.~Feretti,
  %``The Origin and Evolution of Cosmic Magnetism,''
  New Astron.\ Rev.\  {\bf 48}, 1003 (2004)
  [arXiv:astro-ph/0409100].
  %%CITATION = ASTRE,48,1003;%%

\bb{mpc2}
F.~Govoni and L.~Feretti,
  %``Magnetic Field in Clusters of Galaxies,''
  Int.\ J.\ Mod.\ Phys.\  D {\bf 13}, 1549 (2004)
  [arXiv:astro-ph/0410182].
  %%CITATION = IMPAE,D13,1549;%%

\bb{it}
  F.~Tavecchio, G.~Ghisellini, L.~Foschini, G.~Bonnoli, G.~Ghirlanda and P.~Coppi,
  %``The intergalactic magnetic field constrained by Fermi/LAT observations of
  %the TeV blazar 1ES 0229+200,''
  arXiv:1004.1329 [astro-ph.CO].
  %%CITATION = ARXIV:1004.1329;%%

\bb{ru}
  A.~Neronov and I.~Vovk,
  %``Evidence for strong extragalactic magnetic fields from Fermi observations
  %of TeV blazars,''
  Science {\bf 328}, 73 (2010)
  [arXiv:1006.3504 [astro-ph.HE]].
  %%CITATION = SCIEA,328,73;%%

\bb{Forbes:2000gr}
  M.~M.~Forbes and A.~R.~Zhitnitsky,
  %``Primordial galactic magnetic fields from domain walls at the QCD phase
  %transition,''
  Phys.\ Rev.\ Lett.\  {\bf 85}, 5268 (2000)
  [arXiv:hep-ph/0004051].
  %%CITATION = PRLTA,85,5268;%%
  %\cite{Forbes:2001vya}

\bb{Forbes:2001vya}
  M.~M.~Forbes and A.~R.~Zhitnitsky,
  %``Primordial galactic magnetic fields: An application of QCD domain  walls,''
  arXiv:hep-ph/0102158.
  %%CITATION = HEP-PH/0102158;%%
  
\bb{optim}
R.~Banerjee and K.~Jedamzik,
  %``The Evolution of Cosmic Magnetic Fields: From the Very Early Universe, to
  %Recombination, to the Present,''
  Phys.\ Rev.\  D {\bf 70}, 123003 (2004)
  [arXiv:astro-ph/0410032].
  %%CITATION = PHRVA,D70,123003;%%

\bb{chiara1}
C.~Caprini, R.~Durrer and E.~Fenu,
  %``Can the observed large scale magnetic fields be seeded by helical
  %primordial fields?,''
  arXiv:0906.4976 [astro-ph.CO].
  %%CITATION = ARXIV:0906.4976;%%

\bb{chiara2}
C.~Caprini, R.~Durrer and G.~Servant,
  %``The stochastic gravitational wave background from turbulence and magnetic
  %fields generated by a first-order phase transition,''
  arXiv:0909.0622 [astro-ph.CO].
  %%CITATION = ARXIV:0909.0622;%%

\bb{super}
P.~P.~Kronberg,
  %``Extragalactic radio sources, IGM magnetic fields, and AGN-based energy flows,''
  Astron.\ Nachr.\ {\bf 327}, 517 (2006).

\bb{hyper}
Y.~Xu, P.~P.~Kronberg, S.~Habib and Q.~W.~Dufton,
  %``A Faraday Rotation Search for Magnetic Fields in Large Scale Structure,''
  Astrophys.\ J.\  {\bf 637}, 19 (2006)
  [arXiv:astro-ph/0509826].
  %%CITATION = ASJOA,637,19;%%

\bb{eearly}
P.~P.~Kronberg, M.~L.~Bernet, F.~Miniati, S.~J.~Lilly, M.~B.~Short and D.~M.~Higdon,
  %``A Global Probe of Cosmic Magnetic Fields to High Redshifts,''
  Astrophys.\ J.\  {\bf 676}, 7079 (2008)
  [arXiv:0712.0435 [astro-ph]].
  %%CITATION = ASJOA,676,7079;%%

\bb{early}
M.~L.~Bernet, F.~Miniati, S.~J.~Lilly, P.~P.~Kronberg and M.~Dessauges-Zavadsky,
  %``Strong magnetic fields in normal galaxies at high redshifts,''
  arXiv:0807.3347 [astro-ph].
  %%CITATION = ARXIV:0807.3347;%%

\bb{bbn}
G.~Steigman,
  %``Primordial Nucleosynthesis in the Precision Cosmology Era,''
  Ann.\ Rev.\ Nucl.\ Part.\ Sci.\  {\bf 57}, 463 (2007)
  [arXiv:0712.1100 [astro-ph]].
  %%CITATION = ARNUA,57,463;%%

\bb{gw1}
C.~Caprini and R.~Durrer,
  %``Gravitational wave production: A strong constraint on primordial  magnetic
  %fields,''
  Phys.\ Rev.\  D {\bf 65}, 023517 (2001)
  [arXiv:astro-ph/0106244].
  %%CITATION = PHRVA,D65,023517;%%

\bb{gw2}
C.~Caprini and R.~Durrer,
  %``Gravitational waves from stochastic relativistic sources: Primordial
  %turbulence and magnetic fields,''
  Phys.\ Rev.\  D {\bf 74}, 063521 (2006)
  [arXiv:astro-ph/0603476].
  %%CITATION = PHRVA,D74,063521;%%

\bb{slava}
V.~Demozzi, V.~Mukhanov and H.~Rubinstein,
  %``Magnetic fields from inflation?,''
  arXiv:0907.1030 [astro-ph.CO].
  %%CITATION = ARXIV:0907.1030;%%

\bb{coles}
P.~Coles,
  Comments Astroph.\  {\bf 16}, 45 (1992)

\bb{kim}
E.~j.~Kim, A.~Olinto and R.~Rosner,
  %``Generation of density perturbations by primordial magnetic fields,''
  Astrophys.\ J.\  {\bf 468}, 28 (1996)
  [arXiv:astro-ph/9412070].
  %%CITATION = ASJOA,468,28;%%

\bb{fila1}
E.~Battaner, E.~Florido and J.~Jimenez-Vicente,
  %``Magnetic Fields and Large Scale Structure in a radiation dominated
  %Universe,''
  Astron.\ Astrophys.\  {\bf 326}, 13 (1997)
  [arXiv:astro-ph/9602097].
  %%CITATION = AAEJA,326,13;%%

\bb{fila2}
E.~Battaner, E.~Florido and J.~M.~Garcia-Ruiz,
  %``Magnetic fields and large scale structure in a hot Universe. III. The
  %polyhedric network,''
  Astron.\ Astrophys.\  {\bf 327}, 8 (1997)
  [arXiv:astro-ph/9710074].
  %%CITATION = AAEJA,327,8;%%

\bb{fila3}
E.~Florido and E.~Battaner,
  Astron.\ Astrophys.\  {\bf 327}, 1 (1997)

\bb{totani}
T.~Totani,
  Astrophys.\ J.\  {\bf 517}, L69 (1999)

\bb{de1}
A.~G.~Riess {\it et al.}  [Supernova Search Team Collaboration],
  %``Observational Evidence from Supernovae for an Accelerating Universe and a
  %Cosmological Constant,''
  Astron.\ J.\  {\bf 116}, 1009 (1998)
  [arXiv:astro-ph/9805201].
  %%CITATION = ANJOA,116,1009;%%

\bb{de2}
S.~Perlmutter {\it et al.}  [Supernova Cosmology Project Collaboration],
  %``Measurements of Omega and Lambda from 42 High-Redshift Supernovae,''
  Astrophys.\ J.\  {\bf 517}, 565 (1999)
  [arXiv:astro-ph/9812133].
  %%CITATION = ASJOA,517,565;%%

\bb{wmap}
  E.~Komatsu {\it et al.}  [WMAP Collaboration],
  %``Five-Year Wilkinson Microwave Anisotropy Probe (WMAP\altaffilmark 1 )
  %Observations:Cosmological Interpretation,''
  Astrophys.\ J.\ Suppl.\  {\bf 180}, 330 (2009)
  [arXiv:0803.0547 [astro-ph]].
  %%CITATION = APJSA,180,330;%%

\bb{dyn}
F.~R.~Urban and A.~R.~Zhitnitsky,
  %``The QCD nature of Dark Energy,''
  Nucl.\ Phys.\  B {\bf 835}, 135 (2010)
  [arXiv:0909.2684 [astro-ph.CO]].
  %%CITATION = NUPHA,B835,135;%%

\bb{4d}
F.~R.~Urban and A.~R.~Zhitnitsky,
  %``The cosmological constant from the QCD Veneziano ghost,''
  Phys.\ Lett.\  B {\bf 688}, 9 (2010)
  [arXiv:0906.2162 [gr-qc]].
  %%CITATION = PHLTA,B688,9;%%

\bb{ven}
G.~Veneziano,
  %``U(1) Without Instantons,''
  Nucl.\ Phys.\  B {\bf 159}, 213 (1979).
  %%CITATION = NUPHA,B159,213;%%

\bb{witten}
  E.~Witten,
  %``Current Algebra Theorems For The U(1) Goldstone Boson,''
  Nucl.\ Phys.\  B {\bf 156}, 269 (1979).
  %%CITATION = NUPHA,B156,269;%%

\bb{eft}
E.~C.~Thomas, F.~R.~Urban and A.~R.~Zhitnitsky,
  %``The cosmological constant as a manifestation of the conformal anomaly?,''
  JHEP {\bf 0908}, 043 (2009)
  [arXiv:0904.3779 [gr-qc]].
  %%CITATION = JHEPA,0908,043;%%

\bb{2d}
F.~R.~Urban and A.~R.~Zhitnitsky,
  %``The cosmological constant from the ghost. A toy model,''
  Phys.\ Rev.\ D {\bf 80}, 063001 (2009)
  [arXiv:0906.2165 [hep-th]].
  %%CITATION = ARXIV:0906.2165;%%

\bb{cmb}
F.~R.~Urban and A.~R.~Zhitnitsky,
  %``Cosmological constant, violation of cosmological isotropy and CMB,''
  JCAP {\bf 0909}, 018 (2009)
  [arXiv:0906.3546 [astro-ph.CO]].
  %%CITATION = JCAPA,0909,018;%%

\bb{l1}
P.~Di Vecchia and G.~Veneziano,
  %``Chiral Dynamics In The Large N Limit,''
  Nucl.\ Phys.\  B {\bf 171}, 253 (1980).
  %%CITATION = NUPHA,B171,253;%%

\bb{l2}
C.~Rosenzweig, J.~Schechter and C.~G.~Trahern,
  %``Is The Effective Lagrangian For QCD A Sigma Model?,''
  Phys.\ Rev.\  D {\bf 21}, 3388 (1980).
  %%CITATION = PHRVA,D21,3388;%%


\bb{huang}
K.~Huang,  ``Quarks, Leptons And Gauge Fields,'' {\it  World Scientific}, Singapore, Singapore (1982)


\bb{coup1}
P.~Sikivie,
  %``Experimental tests of the *invisible* axion,''
  Phys.\ Rev.\ Lett.\  {\bf 51}, 1415 (1983)
  [Erratum-ibid.\  {\bf 52}, 695 (1984)].
  %%CITATION = PRLTA,51,1415;%%

\bb{coup2}
G.~Raffelt and L.~Stodolsky,
  %``Mixing of the Photon with Low Mass Particles,''
  Phys.\ Rev.\  D {\bf 37}, 1237 (1988).
  %%CITATION = PHRVA,D37,1237;%%

\bb{ps1}
M.~S.~Turner and L.~M.~Widrow,
  %``Inflation Produced, Large Scale Magnetic Fields,''
  Phys.\ Rev.\  D {\bf 37}, 2743 (1988).
  %%CITATION = PHRVA,D37,2743;%%

\bb{ps2}
S.~M.~Carroll and G.~B.~Field,
  %``The Einstein equivalence principle and the polarization of radio
  %galaxies,''
  Phys.\ Rev.\  D {\bf 43}, 3789 (1991).
  %%CITATION = PHRVA,D43,3789;%%

\bb{coup3}
D.~Harari and P.~Sikivie,
  %``Effects of a Nambu-Goldstone boson on the polarization of radio galaxies
  %and the cosmic microwave background,''
  Phys.\ Lett.\  B {\bf 289}, 67 (1992).
  %%CITATION = PHLTA,B289,67;%%

\bb{ps3}
W.~D.~Garretson, G.~B.~Field and S.~M.~Carroll,
  %``Primordial magnetic fields from pseudoGoldstone bosons,''
  Phys.\ Rev.\  D {\bf 46}, 5346 (1992)
  [arXiv:hep-ph/9209238].
  %%CITATION = PHRVA,D46,5346;%%

\bb{coup4}
E.~D.~Carlson and W.~D.~Garretson,
  %``Photon To Pseudoscalar Conversion In The Interstellar Medium,''
  Phys.\ Lett.\  B {\bf 336}, 431 (1994).
  %%CITATION = PHLTA,B336,431;%%

\bb{ps4}
R.~Brustein and D.~H.~Oaknin,
  %``Electroweak baryogenesis induced by a scalar field,''
  Phys.\ Rev.\ Lett.\  {\bf 82}, 2628 (1999)
  [arXiv:hep-ph/9809365].
  %%CITATION = PRLTA,82,2628;%%

\bb{ps5}
R.~Brustein and D.~H.~Oaknin,
  %``Amplification of hypercharge electromagnetic fields by a cosmological
  %pseudoscalar,''
  Phys.\ Rev.\  D {\bf 60}, 023508 (1999)
  [arXiv:hep-ph/9901242].
  %%CITATION = PHRVA,D60,023508;%%

\bb{vanB}
 % \bibitem{Asztalos:2006kz}
  S.~J.~Asztalos, L.~J.~Rosenberg, K.~van Bibber, P.~Sikivie and K.~Zioutas,
  %``Searches for astrophysical and cosmological axions,''
  Ann.\ Rev.\ Nucl.\ Part.\ Sci.\  {\bf 56}, 293 (2006).

\bb{vanBibber:2006rb}
  K.~van Bibber and L.~J.~Rosenberg,
  %``Ultrasensitive Searches For The Axion,''
  Phys.\ Today {\bf 59N8}, 30 (2006).

\bb{eric}
A.~R.~Zhitnitsky,
  %``The Gauge Fields and Ghosts in Rindler Space,''
  arXiv:1004.2040 [gr-qc].
  %%CITATION = ARXIV:1004.2040;%%

\bb{Halperin:1998rc}
  I.~E.~Halperin and A.~Zhitnitsky,
  %``Anomalous effective Lagrangian and theta dependence in {QCD} at finite
  %N(c),''
  Phys.\ Rev.\ Lett.\  {\bf 81}, 4071 (1998)
  [arXiv:hep-ph/9803301].
  %%CITATION = PRLTA,81,4071;%%

\bb{Kim:2010gf}
  J.~Kim and P.~Naselsky,
  %``Anomalous parity asymmetry of the Wilkinson Microwave Anisotropy Probe
  %power spectrum data at low multipoles,''
  Astrophys.\ J.\  {\bf 714}, L265 (2010)
  [arXiv:1001.4613 [Unknown]].
  %%CITATION = ASJOA,714,L265;%%

\bb{Kim:2010gd}
  J.~Kim and P.~Naselsky,
  %``Anomalous parity asymmetry of WMAP power spectrum data at low multpoles: is
  %it cosmological or systematics?,''
  arXiv:1002.0148 [Unknown].
  %%CITATION = ARXIV:1002.0148;%%

\bb{hel1}
L.~Pogosian, T.~Vachaspati and S.~Winitzki,
  %``Signatures of kinetic and magnetic helicity in the CMBR,''
  Phys.\ Rev.\  D {\bf 65}, 083502 (2002)
  [arXiv:astro-ph/0112536].
  %%CITATION = PHRVA,D65,083502;%%

\bb{hel2}
T.~Kahniashvili and B.~Ratra,
  %``Effects of Cosmological Magnetic Helicity on the Cosmic Microwave
  %Background,''
  Phys.\ Rev.\  D {\bf 71}, 103006 (2005)
  [arXiv:astro-ph/0503709].
  %%CITATION = PHRVA,D71,103006;%%

\bb{pol1}
A.~Kosowsky and A.~Loeb,
  %``Faraday Rotation of Microwave Background Polarization by a Primordial
  %Magnetic Field,''
  Astrophys.\ J.\  {\bf 469}, 1 (1996)
  [arXiv:astro-ph/9601055].
  %%CITATION = ASJOA,469,1;%%

\bb{pol2}
J.~A.~Adams, U.~H.~Danielsson, D.~Grasso and H.~Rubinstein,
  %``Distortion of the acoustic peaks in the CMBR due to a primordial magnetic
  %field,''
  Phys.\ Lett.\  B {\bf 388}, 253 (1996)
  [arXiv:astro-ph/9607043].
  %%CITATION = PHLTA,B388,253;%%

\bb{pol3}
D.~D.~Harari, J.~D.~Hayward and M.~Zaldarriaga,
  %``Depolarization of the cosmic microwave background by a primordial magnetic
  %field and its effect upon temperature anisotropy,''
  Phys.\ Rev.\  D {\bf 55}, 1841 (1997)
  [arXiv:astro-ph/9608098].
  %%CITATION = PHRVA,D55,1841;%%

\bb{pol4}
M.~Giovannini,
  %``Cosmic microwave background polarization, Faraday rotation, and  stochastic
  %gravity-waves backgrounds,''
  Phys.\ Rev.\  D {\bf 56}, 3198 (1997)
  [arXiv:hep-th/9706201].
  %%CITATION = PHRVA,D56,3198;%%

\bb{pol5}
E.~S.~Scannapieco and P.~G.~Ferreira,
  %``Polarization - temperature correlation from primordial magnetic field,''
  Phys.\ Rev.\  D {\bf 56}, 7493 (1997)
  [arXiv:astro-ph/9707115].
  %%CITATION = PHRVA,D56,7493;%%

\bb{gk1}
M.~Giovannini and K.~E.~Kunze,
  %``A magnetized completion of the $\Lambda$CDM paradigm,''
  Phys.\ Rev.\  D {\bf 77}, 061301 (2008)
  [arXiv:0712.1977 [astro-ph]].
  %%CITATION = PHRVA,D77,061301;%%

\bb{gk2}
M.~Giovannini and K.~E.~Kunze,
  %``Magnetized CMB observables: a dedicated numerical approach,''
  Phys.\ Rev.\  D {\bf 77}, 063003 (2008)
  [arXiv:0712.3483 [astro-ph]].
  %%CITATION = PHRVA,D77,063003;%%

\bb{gk3}
M.~Giovannini and K.~E.~Kunze,
  %``Generalized CMB initial conditions with pre-equality magnetic fields,''
  Phys.\ Rev.\  D {\bf 77}, 123001 (2008)
  [arXiv:0802.1053 [astro-ph]].
  %%CITATION = PHRVA,D77,123001;%%

\bb{gk4}
M.~Giovannini and K.~E.~Kunze,
  %``Faraday rotation, stochastic magnetic fields and CMB maps,''
  Phys.\ Rev.\  D {\bf 78}, 023010 (2008)
  [arXiv:0804.3380 [astro-ph]].
  %%CITATION = PHRVA,D78,023010;%%

\bb{gk5}
M.~Giovannini and K.~E.~Kunze,
  %``Cosmic polarimetry in magnetoactive plasmas,''
  Phys.\ Rev.\  D {\bf 79}, 063007 (2009)
  [arXiv:0812.2207 [astro-ph]].
  %%CITATION = PHRVA,D79,063007;%%
\exclude{
\bb{pav1}
J.~Kim and P.~Naselsky,
  %``Anomalous parity asymmetry of WMAP power spectrum data at low multpoles: is
  %it cosmological or systematics?,''
  arXiv:1002.0148 [astro-ph.CO].
  %%CITATION = ARXIV:1002.0148;%%

\bb{pav2}
J.~Kim and P.~Naselsky,
  %``Anomalous parity asymmetry of the WMAP power spectrum data at low
  %multipoles,''
  arXiv:1001.4613 [astro-ph.CO].
  %%CITATION = ARXIV:1001.4613;%%
}
\bb{bir1}
J.~N.~Clarke, G.~Karl and P.~J.~S.~Watson,
  %``The Optical Activity Of Intergalactic Space,''
  Can.\ J.\ Phys.\  {\bf 60}, 1561 (1982).
  %%CITATION = CJPHA,60,1561;%%

\bb{bir2}
M.~Giovannini,
  %``Magnetized birefringence and CMB polarization,''
  Phys.\ Rev.\  D {\bf 71}, 021301 (2005)
  [arXiv:hep-ph/0410387].
  %%CITATION = PHRVA,D71,021301;%%

\bb{bir3}
M.~Giovannini and K.~E.~Kunze,
  %``Birefringence, CMB polarization and magnetized B-mode,''
  Phys.\ Rev.\  D {\bf 79}, 087301 (2009)
  [arXiv:0812.2804 [astro-ph]].
  %%CITATION = PHRVA,D79,087301;%%

\bb{pv1}
S.~M.~Carroll, G.~B.~Field and R.~Jackiw,
  %``Limits on a Lorentz and Parity Violating Modification of Electrodynamics,''
  Phys.\ Rev.\  D {\bf 41}, 1231 (1990).
  %%CITATION = PHRVA,D41,1231;%%

\bb{pv2}
N.~F.~Lepora,
  %``Cosmological Birefringence and the Microwave Background,''
  arXiv:gr-qc/9812077.
  %%CITATION = GR-QC/9812077;%%

\bb{pv3}
A.~Lue, L.~M.~Wang and M.~Kamionkowski,
  %``Cosmological signature of new parity-violating interactions,''
  Phys.\ Rev.\ Lett.\  {\bf 83}, 1506 (1999)
  [arXiv:astro-ph/9812088].
  %%CITATION = PRLTA,83,1506;%%

\bb{pv4}
K.~R.~S.~Balaji, R.~H.~Brandenberger and D.~A.~Easson,
  %``Spectral dependence of CMB polarization and parity,''
  JCAP {\bf 0312}, 008 (2003)
  [arXiv:hep-ph/0310368].
  %%CITATION = JCAPA,0312,008;%%

\bb{hut1}
D.~Hutsemekers,
  Astron.\ Astrophys.\  {\bf 332}, 410 (1998)

\bb{hut2}
D.~Hutsemekers and H.~Lamy,
  %``Confirmation of the existence of coherent orientations of quasar
  %polarization vectors on cosmological scales,''
  Astron.\ Astrophys.\  {\bf 367}, 381 (2001)
  [arXiv:astro-ph/0012182].
  %%CITATION = ASTRO-PH/0012182;%%

\bb{hut3}
D.~Hutsemekers, R.~Cabanac, H.~Lamy and D.~Sluse,
  %``Mapping extreme-scale alignments of quasar polarization vectors,''
  Astron.\ Astrophys.\  {\bf 441}, 915 (2005)
  [arXiv:astro-ph/0507274].
  %%CITATION = AAEJA,441,915;%%

\bb{pa1}
S.~A.~Joshi, R.~A.~Battye, I.~W.~A.~Browne, N.~Jackson, T.~W.~B.~Muxlow and P.~N.~Wilkinson,
  %``The polarization in the JVAS/CLASS flat-spectrum radio sources: II. A
  %search for aligned radio polarizations,''
  Mon.\ Not.\ Roy.\ Astron.\ Soc.\  {\bf 380}, 162 (2007)
  [arXiv:0705.2548 [astro-ph]].
  %%CITATION = MNRAA,380,162;%%

\bb{pa2}
R.~A.~Battye, I.~W.~A.~Browne and N.~Jackson,
  %``Biases in the polarization position angles in the NVSS point source
  %catalogue,''
  Mon.\ Not.\ Roy.\ Astron.\ Soc.\  {\bf 85}, 274 (2008)
  [arXiv:0902.1619 [astro-ph.CO]].
  %%CITATION = MNRAA,85,274;%%

\bb{jain1}
S.~Das, P.~Jain, J.~P.~Ralston and R.~Saha,
  %``Probing dark energy with light: Propagation and spontaneous
  %polarization,''
  JCAP {\bf 0506}, 002 (2005)
  [arXiv:hep-ph/0408198].
  %%CITATION = JCAPA,0506,002;%%

\bb{jain2}
S.~Das, P.~Jain, J.~P.~Ralston and R.~Saha,
  %``The dynamical mixing of light and pseudoscalar fields,''
  Pramana {\bf 70}, 439 (2008)
  [arXiv:hep-ph/0410006].
  %%CITATION = PRAMC,70,439;%%

\bb{jain3}
A.~K.~Ganguly, P.~Jain and S.~Mandal,
  %``Photon and axion oscillation in a magnetized medium: A general treatment,''
  Phys.\ Rev.\  D {\bf 79}, 115014 (2009)
  [arXiv:0810.4380 [hep-ph]].
  %%CITATION = PHRVA,D79,115014;%%

\bb{Carroll:1998zi}
  S.~M.~Carroll,
  %``Quintessence and the rest of the world,''
  Phys.\ Rev.\ Lett.\  {\bf 81}, 3067 (1998)
  [arXiv:astro-ph/9806099].
  %%CITATION = PRLTA,81,3067;%%
  
\end{thebibliography}
\end{document}